# Flow Induced Corrosion in 6061-T6 Aluminum Pipes in De-Ionized Water Environments

Abhishek Deshpande[†] and Erik Voirin
Fermi National Accelerator Laboratory
Kirk Rd. and Pine St.
Batavia, Illinois 60510
USA

## ABSTRACT

Flow induced corrosion/erosion of 6061-T6 aluminum in de-ionized (DI) water environments has not been studied widely. Especially, the long-term effects of corrosion/erosion in seemingly benign flow velocity, temperature, and resistivity ranges of 8 ft/s, 85 ºF, 3-5 MOhm-cm, respectively. This study concludes that the flow induced corrosion/erosion in the above parameter ranges is minimal. This is detailed by presenting a literature survey, measuring pipe wall samples from a system that has operated in the above parameter range, predict the loss of material in mm/year at the above velocities and at higher temperatures using electric potential values from other experimental studies coupled with wall shear stress simulated using Computational Fluid Dynamic (CFD) analysis.

Key words: De-ionized water, 6061-T6 Aluminum, Flow Induced Corrosion/Erosion

## INTRODUCTION

The LBNF Absorber, the component that absorbs the residual proton beam, consists of 6061-T6 aluminum core blocks. The core blocks are water cooled with de-ionized (DI) water which becomes radioactive during beam operations. Thermo-mechanical simulations have guided engineering design to set the fluid velocity in the core block's gun-drilled channels to approximately 8 ft/s. The water inlet temperature is 80 ºF. The maximum water outlet temperature on any of the core blocks does not exceed 85 ºF. The system's de-ionization loop will be maintained between 3-5 MOhm-cm.

This study includes a calculation that uses parameters from research articles to predict the loss of material in mm/year at the above velocities and resistivity values and at higher temperatures. Also, this study presents a prediction of maximum wall shear stress using CFD analysis. It describes operational experience of operating the Main Ring Low Conductivity Water (LCW) cooling system, which has 6061-T6 supply and return headers, for over 40-years. Sections of this piping system are still in operation and support a portion of the Fermilab's accelerator complex. Also, conclusions from a literature survey on the effect of high velocity DI water flows on aluminum components are discussed.

## LITERATURE SURVEY

A corrosion study was done at Savanah River where aluminum samples were immersed in DI water for 85-months.[1] Metallographic analysis was done on the samples after this time. The samples were pure 6061 and welded versions of the same. Filler metal used for the welded samples was R4043. The DI level and the pH of the water were maintained at 0.1 MOhm-cm and 5.5-8.5, respectively. The samples exhibited minimal uniform corrosion. The unwelded and welded samples experienced a weight gain of approximately 1.1% and 1.4%, respectively. [1] The unwelded 6061 samples displayed no pitting corrosion. The welded versions on the other hand revealed pitting underneath the oxide layer in the weld metal. However, the maximum depth of pitting was approximately 0.005 in.

A thorough literature survey of the effect of erosion or corrosion in aluminum components subjected to high DI water velocities was conducted. This body of literature is sparse. Erosion tests on 2S and 63S aluminum components which had small, machined openings were performed.[2] Filtered de-ionized water at various velocities was pumped through these components and were tested for erosion and corrosion. It was concluded that both alloys were entirely resistant to frictional erosion at fluid velocities of up to 50 ft/s. The fluid's nominal temperature, pressure, and resistivity were 86 ºF, 355 Psig, and 0.5 MOhm-cm, respectively. It must be noted that the composition of aluminum alloys used in this study is different from 6061-T6. However, the aluminum content in these alloys is almost equivalent to that in 6061-T6. Table 1 below highlights this. The oxide layer protects the underlying metal from erosion corrosion, thus presumably the quality of this layer for 6061-T6 is equivalent to that of 63S and 2S and would offer comparable resistance to wear.

**Table 1: Comparison of alloying element and aluminum percentages in 6061-T6, 63S and 2S aluminum alloys.**

|  | Alloying Elements | | | | | |
| ---: | ---: | ---: | ---: | ---: | ---: | ---: |
| Aluminum type | Si | Mg | Cu | Cr | Fe | Al |
| 6061-T6 | 0.6% | 1% | 0.25% | 0.2% | 0% | 97.95% |
| 63S | 0.4% | 0.7% | 0.04% | 0% | 0.2% | 98.66% |
| 2S | 0.2% | 0.01% | 0.03% | 0% | 0.3% | 99.46% |
| %difference 6061 and 63S | 40.00% | 35.29% | 144.83% | 200% | 200% | 0.72% |
| %difference 6061 and 2S | 100.00% | 196.04% | 157.14% | 200% | 200% | 1.53% |

## RESULTS AND DISCUSSION

**Operational Experience: Main Ring Low Conductivity Water (LCW) System**

The Main Ring LCW system consisted of 24 pump houses supplying cooling water to the Main Ring particle accelerator. The system was commissioned in the mid *'70s*. Sections of the cooling system continue to support a portion of the accelerator complex to this day. The main supply and return headers for the system are 5" NPS 6061-T6 ANSI B241 Sch40 pipes. The average flow in these headers was 375 Gpm. This corresponds to ~ 6 ft/s. The system lacks full-flow or side stream filtration. The average operation temperature, pressure, and resistivity are 90 °F, 160 Psig, and 10 MOhm-cm, respectively. Samples shown in Fig 1 were cut out from a decommissioned header's 90-degree elbow of the system. This section of the header operated for over 40-years at the above conditions. Sample thicknesses are highlighted in Table 2:

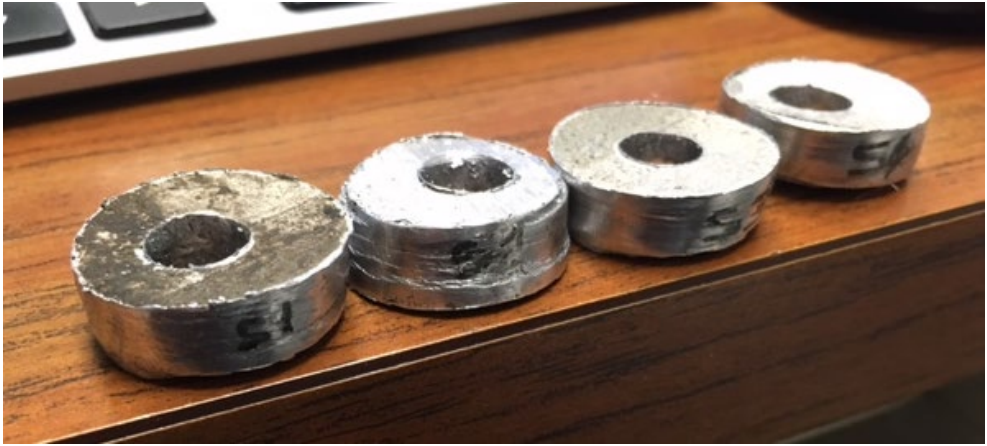

**Fig 1: Samples cut from the supply header of the Main Ring LCW cooling system.**

**Table 2: Sample thickness compared to the nominal wall thickness of 5" NPS Sch40 pipe.**

| Sample ID | Min thickness, in | Nominal Sch40 pipe wall thickness, in. | Material loss or gain, in. |
|---|---|---|---|
| S1 | 0.254 | 0.258 | -0.0040 |
| S2 | 0.2545 | 0.258 | -0.0035 |
| S3 | 0.2515 | 0.258 | -0.0065 |
| S4 | 0.262 | 0.258 | 0.0040 |

The above table shows that the wall thickness has changed minimally over 40-years of operations.

**Estimation of Material Loss**

A CFD simulation was performed to determine the wall shear stress on the miter elbows, which are present on the inlet and outlet ports of each core block, at water velocity and temperature of 8 ft/s and 130 °F, respectively. A maximum wall shear stress of 40 Pa was simulated for a condition where flow exits the elbow from the top, Fig 2. Flow at the inlet has a maximum shear stress of 50 Pa, however, since the water at the outlet would be warmer, this is considered the worst case. Investigating why wall shear stress would be indicative of corrosion, literature sources suggest high shear stresses bring more of the corrosive components contained in the fluid through the fluid boundary layer and into contact with the wall. The physics behind corrosion/erosion may be understood as a convective mass transfer coefficient at the fluid solid interface, which is analogous to a convective heat transfer coefficient, acting alongside shear stress. It may be theorized that mass transfer is the variable of interest for corrosion, not purely wall shear stress, and this is validated with references.[3] Fick's equations for mass transfer and the Fourier heat equation are analogous differential equations and of the same form. Heat and mass transfer are captured by the Lewis number, which is the ratio of Schmidt number over the Prandtl number. Both heat and mass transfer can be simulated using commercial computational dynamics code, however, in this study, only shear stress is compared as it seems to be the more common variable used in corrosion studies.

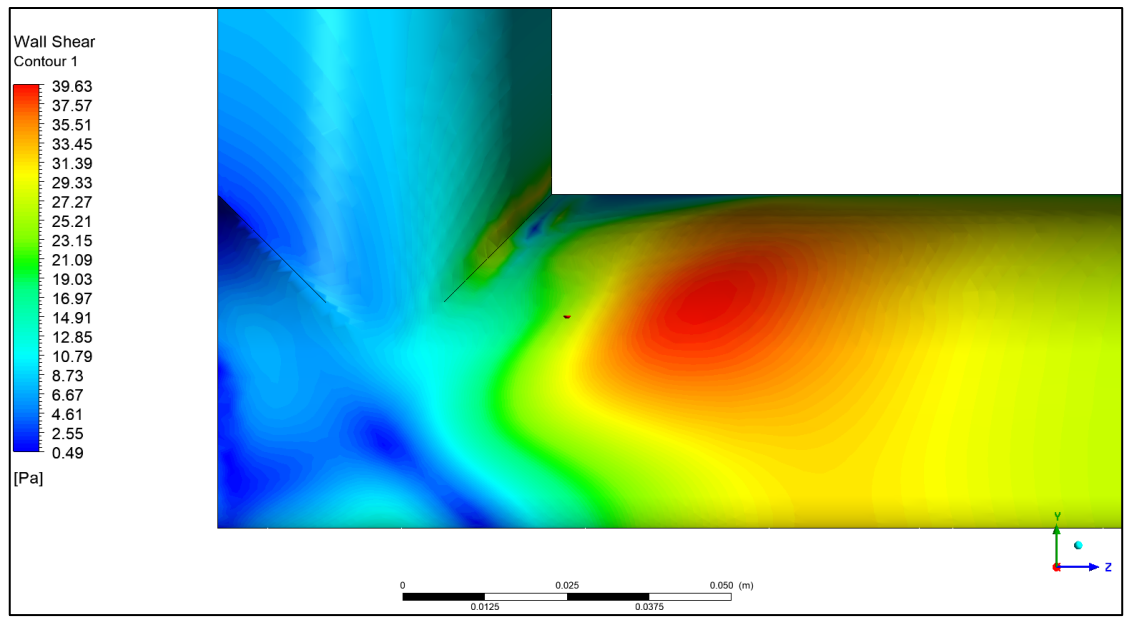

**Fig 2: Shear stress in elbow at outlet of cooling block, where water will be hotter, (with more corrosion potential.) Water is flowing from the top, down then to the right in this image.**

Literature survey revealed two sources which had measurements of aluminum corrosion in flow scenarios. In the first, an impinging jet directed onto an aluminum disc at multiple angles of attack was studied.[4] The water velocity was 3 m/sec (9.8 ft/sec). This is test case #1. Shear stresses at the multiple angles of attack spanned the shear stresses seen in the CFD simulation. The relative mass transfer coefficient and corrosion potential vs. shear stress for this study should be higher than our case for several reasons: First, sand was entrained in the fluid, meaning erosion-corrosion was contributing to corrosion. Second, the impinging jet onto the aluminum disc enables all corrosive components in the fluid to convect across the disc surface, whereas in an aluminum tube, these components will interact with a much larger aluminum surface prior to interacting with the area of interest, reducing boundary layer concentration as well as the local mass transfer.

In the second study, DI water was flowing past aluminum electrodes, again at 3 m/s (9.8 ft/sec) mean velocity.[5] Shear stress measurements were not performed here, and there was not enough detail on the geometry to determine this value. This is test case #2. However, analysis of our CFD model's local mass transfer coefficients show that, the peak value is less than 3 times that in plain straight tube before the elbow. This implies that the miter elbow geometry does not increase the mass transfer coefficient appreciably. Hence likely contributing to corrosion/erosion rates to the same degree or less as in a straight tube. Of the two test cases discussed, test case #2 matches closely with the miter elbow geometry.

Calculations of corrosion rates using data from test cases #1 and #2 are shown below. Test case #1 shows a higher corrosion rate than test case #2, which is likely due to the erosion-corrosion contribution of the sand particles in test case #1. Both cases show low corrosion rates and imply a maximum 35-year material loss on the order of 2.22 mm (0.087").

$$I_{current} = \frac{V}{R} \quad (1)$$

$$W_{lost} = \frac{I_{current} M_{Al}}{n_{elect} F_c} \quad (2)$$

$$L_{rate} = \frac{W_{Lost}}{\rho_{Al}} \quad (3)$$

$$T_{lost} = L_{rate} 35\, y \quad (4)$$

Where,

| | | |
|---|---|---|
| $I_{current}$ | = | Corrosion current density, µA/cm² |
| $V$ | = | Corrosion potential, 0.7 V for test case 1 and 0.93 for test case #2 |
| $R$ | = | Charge transfer impedance, 120000 Ohm cm² for test case #1 |
| $W_{lost}$ | = | Corrosion rate, Kg/s m² |
| $M_{Al}$ | = | Molecular mass of aluminum, 27 g/mol |
| $n_{elect}$ | = | Valance for aluminum, 3 [6] |
| $F_c$ | = | Faraday constant, 96500 C/mol [6] |
| $L_{rate}$ | = | Loss rate per year, mm/yr |
| $\rho_{Al}$ | = | Density of aluminum, 2700 Kg/m³ |
| $T_{lost}$ | = | Total expected material lost during 35 years of operation |

**Table 4.1: Corrosion parameters and material loss summary.**

| Parameter | Test case #1 | Test case #2 | Units |
|---|---|---|---|
| $I_{current}$ | 5.83 | 1.11 | µA/cm² |
| $W_{lost}$ | 5.44E-09 | 1.04E-09 | Kg/s m² |
| $L_{rate}$ | 0.064 | 0.012 | mm/yr |
| $T_{lost}$ | 2.22 | 0.42 | mm |

## CONCLUSIONS

A literature survey on flow induced corrosion of aluminum alloys in de-ionized water environments is sparse. The available experimental studies document minimal loss of material in aluminum alloys at varying flow conditions.[1,2] Samples of aluminum piping that operated for over 40-years in the Main Ring LCW system revealed little to no corrosion damage.

CFD simulations revealed that the worst-case shear stress on the aluminum components of the LBNF absorber is very low, 50 Pa. Parameters derived from two experimental setups were used to estimate the material loss in aluminum alloys owing to flow induced corrosion/erosion. The loss of material in both cases over 35-years was minimal. The material lost in one case was much higher than in the other. This may be attributed to the entrained sand present in the first case. From the above study one can expect minimal erosion/corrosion in the aluminum components of the LBNF absorber during its ~35-year operational life.

## REFERENCES

[1] P.R. Vormelker and A. J. Duncan., Corrosion Evaluation of Aluminum Alloys in De-ionized Water., WSRC-MS-2004-00654., Savannah River National Laboratory.

[2] F. C. Apple and E. H. Honeycutt., Erosion of Aluminum., May 1957., DP-214, Metallurgy and Ceramics TDI-4500, 13$^{th}$ Ed., Pile Engineering Division., E. I. du Pont de Nemours &Co.

[3] K. G. Jordan and P. R. Rhodes., Corrosion of Carbon Steel by $CO_2$ Solutions: The Role of Fluid Flow., March 1995 Paper No. 95125, Corrosion/95, Orlando, FL.

[4] L.Y. Xu and Y.F. Cheng., Electrochemical Characterization and CFD Simulation of Flow-Assisted Corrosion of Aluminum Alloy in Ethylene Glycol–Water Solution., July 2008., Corrosion Science, Volume 50, Issue 7, Pages 2094-2100.

[5] L. Hao et al., Erosion Corrosion Behavior of Aluminum in Flowing Deionized Water at Various Temperatures. Materials (Basel). February 2020 Feb 8;13(3):779. doi: 10.3390/ma13030779. PMID: 32046276; PMCID: PMC7040843.

[6] M.P. Schultz and G. W. Swain., Corrosion Basics, OCE-4518 Protection of Marine Materials Class Notes, Florida Institute of Technology.